\documentclass[aps,twocolumn,prd,showpacs,nofootinbib]{revtex4}
\usepackage{amsmath}
\usepackage{amssymb}
\usepackage{graphicx}

\newcommand{\lsim}{\lesssim}
\newcommand{\ie}{\emph{i.e.~}}

\newcommand{\eg}{\emph{e.g.~}}

\newcommand{\Mp}{M_\mathrm{Pl}}
\newcommand{\de}[2]{\kern - #1 em \mathrm{d} #2}

\newcommand{\dd}{\mathrm{d}}

\newcommand{\MeV}{\mbox{MeV}}
\newcommand{\GeV}{\mbox{GeV}}
\newcommand{\TeV}{\mbox{TeV}}

\newcommand{\Mpc}{\mbox{Mpc}}
\newcommand{\bmk}{\boldsymbol k}

\newcommand{\ur}{\mathrm{r}}
\newcommand{\urad}{\mathrm{rad}}
\newcommand{\ureh}{\mathrm{reh}}
\newcommand{\uend}{\mathrm{end}}

\newcommand{\unuc}{\mathrm{nuc}}
\newcommand{\ueq}{\mathrm{eq}}

\newcommand{\ub}{\mathrm{b}}
\newcommand{\udm}{\mathrm{dm}}
\newcommand{\ud}{\dd}

\newcommand{\Rrad}{R_\urad}
\newcommand{\Rreh}{R}
\newcommand{\wreh}{\overline{w}_\ureh}
\newcommand{\rhoeq}{\rho_\ueq}
\newcommand{\rhoreh}{\rho_\ureh}
\newcommand{\rhoend}{\rho_\uend}
\newcommand{\radnow}{{\ur_0}}
\newcommand{\rhoradnow}{\rho_{\radnow}}
\newcommand{\rhonuc}{\rho_\unuc}

\newcommand{\Nend}{N_\uend}

\newcommand{\Nstar}{N_*}
\newcommand{\Nzero}{N_0}

\newcommand{\OmegaB}{\Omega_\ub}
\newcommand{\OmegaDM}{\Omega_\udm}

\newcommand{\mdl}{\mathcal{M}}
\newcommand{\params}{\Theta}
\newcommand{\data}{d}
\newcommand{\like}{{\mathcal L}}

\newcommand{\be}{\begin{equation}}
\newcommand{\en}{\end{equation}}
\newcommand{\DKL}{\mathcal{D}_\text{KL}}
\newcommand{\Cb}{\mathcal{C}_\ub}

\begin{document}

\title{Hunting Down the Best Model of Inflation with Bayesian Evidence}

\author{J\'er\^ome Martin} \email{jmartin@iap.fr}
\affiliation{Institut d'Astrophysique de Paris, \\ UMR 7095-CNRS,
Universit\'e Pierre et Marie Curie, \\ 98bis boulevard Arago, 75014
Paris, France}

\author{Christophe Ringeval} \email{christophe.ringeval@uclouvain.be}
\affiliation{Institute of Mathematics and Physics, Centre for
  Cosmology, Particle Physics and Phenomenology, \\ Louvain
  University, 2 Chemin du Cyclotron, 1348 Louvain-la-Neuve, Belgium}

\author{Roberto Trotta} \email{r.trotta@imperial.ac.uk}
\affiliation{Astrophysics Group, Imperial College London, Blackett
  Laboratory, Prince Consort Road, London SW7 2AZ, UK}

\date{\today}

\begin{abstract}
  We present the first calculation of the Bayesian evidence for
  different prototypical single field inflationary scenarios,
  including representative classes of small field and large field
  models. This approach allows us to compare inflationary models in a
  well-defined statistical way and to determine the current ``best
  model of inflation''. The calculation is performed numerically by
  interfacing the inflationary code \texttt{FieldInf} with
  \texttt{MultiNest}. We find that small field models are currently
  preferred, while large field models having a self-interacting
  potential of power $p>4$ are strongly disfavoured. The
  class of small field models as a whole has posterior odds of
  approximately $3:1$ when compared with the large field class. The
  methodology and results presented in this article are an additional
  step toward the construction of a full numerical pipeline to
  constrain the physics of the early Universe with astrophysical
  observations. More accurate data (such as the Planck data) and the
  techniques introduced here should allow us to identify conclusively
  the best inflationary model.
\end{abstract}

\maketitle

\section{Introduction}
\label{sec:introduction}

The theory of inflation represents a cornerstone of the standard model
of modern cosmology~\cite{Guth:1980zm, Linde:1981mu,
  Albrecht:1982wi,Linde:1983gd} (for a review, see \eg
Refs.~\cite{Linde:2007fr, Mukhanov:1990me, Martin:2003bt,
  Martin:2004um, Martin:2007bw}). By definition, it is a phase of
accelerated expansion which is supposed to take place in the very
early universe, somewhere between the electroweak to the Grand Unified
Theory energy scales, \ie between $\sim 10^3\, \GeV$ and $\sim
10^{15}\, \GeV$~\cite{Guo:2010mm}. Inflation allows us to understand
several puzzles which plagued the pre-inflationary standard model and
that could not be understood otherwise. Without inflation, the
standard model of cosmology would remain incomplete and highly
unsatisfactory. The most spectacular achievement of inflation is that,
combined with quantum mechanics, it provides a convincing mechanism
for the origin of the cosmological fluctuations (the seeds of galaxies
and of Cosmic Microwave Background - CMB -
anisotropies)~\cite{Mukhanov:1981xt,Hawking:1982cz,Starobinsky:1982ee,
  Guth:1982ec,Bardeen:1983qw} and it predicts that their spectrum
should be almost scale invariant (\ie equal power on all spatial
scales)~\cite{Stewart:1993bc,Mukhanov:1990me,Liddle:1994dx} which is
fully consistent with the observations~\cite{Martin:2006rs}. This part
of the scenario is particularly remarkable since it combines general
relativity and quantum mechanics.

\par

However, the physical nature of the inflaton (the field driving
inflation) and its relation with the standard model of particle
physics and its extensions remain elusive. Moreover the shape of its
potential is not known except, of course, that it must be sufficiently
flat. This is not so surprising since, as mentioned above, the
inflationary mechanism is supposed to take place at energy scales larger
than typically $\sim 1 \TeV$, in a regime where particle physics is not
known and has not been tested at accelerators.  Another crucial aspect
of the inflationary scenario is how it ends and how it is connected to
the subsequent hot big bang phase. It is believed that, after the
slow-roll period, the field reaches the bottom of its potential,
oscillates and decays into
radiation~\cite{Turner:1983he,Kofman:1997yn,Bassett:2005xm,
  Mazumdar:2010sa}. In this way, inflation is smoothly connected to
the radiation-dominated epoch. However, the energy density at which
the radiation-dominated era starts is not accurately known, although some new
constraints on the reheating have recently been obtained in
Refs.~\cite{Martin:2010kz,Nakayama:2008wy,Kuroyanagi:2009br}.

\par
  
Despite the fact that it has become a cornerstone of modern cosmology,
inflation is not as observationally constrained as the other
components of the standard model. To improve on this situation,
full numerical approaches can be put in place in order to use, in an
optimal way, the astrophysical data now at our disposal
~\cite{Ringeval:2007am, Salopek:1988qh, Grivell:1999wc, Leach:2000yw,
  Adams:2001vc, Makarov:2005uh, Bird:2009pq}. This should allow
investigations on the ``fine structure'' of the inflationary
scenario. This program is particularly timely since new high-accuracy
astrophysical observations, such as the European Space Agency Planck
data~\cite{Lamarre:2003zh}, among others, will be released soon. They
will provide an unprecedented window of opportunity to learn about
inflation.

\par

In this article, we are concerned with the question of how to evaluate
the performance of a given inflationary model to explain the data as
compared with others. This problem can be dealt within Bayesian
inference~\cite{Trotta:2008qt} (see \eg Ref.~\cite{Ballesteros:2007te}
for an application to inflationary model comparison). In fact,
Bayesian statistics can be used at two levels. The first level is to
determine which model parameter values are favoured by the data within
a given inflationary model, and this for all models. To this end, one
needs to compute the model's predictions for the relevant observables,
such as the CMB, the galaxy power spectra, etc., and then use the
experimental data to extract the posterior probability distributions
of the model parameters given the data and the theoretical priors. The
second level is to use Bayesian inference for model comparison. At
this level, one has to calculate, for each model, the global
likelihood (also known as the evidence, or model likelihood) which is
obtained by integrating the usual likelihood over all of the model
parameters' values, weighted by their prior probability
distribution. The resulting quantity can be used to compute the
posterior probability of the model, given the available data, thus
updating our prior belief in each of the inflationary models in light
of the observations. The Bayesian approach to model comparison has the
advantage of automatically incorporating a quantitative notion of
``Occam's razor'', \ie more complex inflationary models are assigned a
larger posterior probability only if their complexity is effectively
required to explain the data.

\par

On the practical side, these two levels in Bayesian inference can be
implemented by adopting appropriate numerical algorithms to integrate
the power spectrum for a given inflationary model. This has been
routinely available for several years now and, in this paper, we use
the public code \texttt{FieldInf}~\cite{Ringeval:2005yn,
  Martin:2006rs, Ringeval:2006}. This inflationary code is then
coupled with a CMB perturbation code, such as
\texttt{CAMB}~\cite{Lewis:1999bs}, and then linked with an appropriate
algorithm capable of delivering both the posterior distributions for
each model's parameters as well as the Bayesian evidence of each
model. The evidence is computed using the
publicly available \texttt{MultiNest}
code~\citep{Feroz:2007kg,Feroz:2008xx,Trotta:2008bp}, which implements
the nested sampling algorithm, employed as an add-on sampler to
\texttt{CosmoMC}~\citep{Lewis:2002ah}.

\par

On the theoretical side, one has to choose classes of scenarios that
are representative of the inflationary landscape and that one wishes
to analyze. In this article, we focus on large and small field models
for reasons specified in the following. The reheating stage is
described via the reheating parameter as introduced in
Refs.~\cite{Martin:2006rs, Martin:2010kz}. Moreover, since the
choice of priors is always relevant in problems of model
comparison, we have paid particular attention to their physical motivation and we
carefully investigate this question both for the parameters describing
the inflationary potential and for the reheating.

\par

This article is organized as follows. In the next section,
Sec.~\ref{sec:infpert}, we present the models studied, paying special
attention to the reheating part and the so-called reheating
parameter. In Sec.~\ref{sec:bayesianevidence}, we recall the
definition of the Bayesian evidence, describing in detail how the
priors on the free parameters characterizing each scenario are
chosen. We also explain how its calculation is implemented
numerically. Finally, in Sec.~\ref{sec:discussion}, we present our
results and discuss their physical implications. Readers already
familiar with the inflationary models, techniques and methods can directly jump to
Sec.~\ref{sec:priors}. Perhaps the most important outcome of our
article is that it sketches a general method which allows us to
quantify and determine the ``best'' model of inflation (within the
list of models considered here).

\section{Inflationary Cosmological Perturbations}
\label{sec:infpert}

In this section, after having briefly recalled how the theory of
cosmological perturbations of quantum-mechanical origin allows us to
derive the inflationary predictions, we present the scenarios studied
here, discuss our choice of parametrization and motivate it based on
physical considerations.

\subsection{Choosing the Inflationary Potential}
\label{subsec:infpot}

In order to compare inflation with various astrophysical observables,
one must first determine the power spectrum of the density
perturbations defined by the following expression:
\begin{equation}
{\cal P}_{\zeta}(k)\equiv \frac{k^3}{2\pi ^2}
\left\vert \zeta_{\bmk}\right\vert ^2
\label{Pzeta}
\end{equation}
where $\zeta_{\bmk}$ is the comoving curvature perturbation in
Fourier space and is a conserved quantity on super-Hubble length
scales~\cite{Mukhanov:1990me,Schwarz:2001vv,
  Martin:2003bt,Martin:2004um,Martin:2007bw}.

This power spectrum depends on the shape of the inflaton's
potential, and thus, on its free parameters which have to be
specified. It is common to describe the landscape of possible single
field inflationary models with three different archetypal classes:
large field models, small field models and hybrid inflation. This
simple approach is based on the following considerations. Any inflaton
potential $V(\phi)$ can always be Taylor expanded as
\begin{equation}
V\left(\phi\right)=V_0\pm \alpha \left(\frac{\phi}{\Mp}\right)^2+\cdots .
\end{equation}
According to the value of the coefficients of the expansion, one
obtains different classes of models. If the constant term $V_0$
vanishes, then one obtains a large field
model~\cite{Linde:1983gd,Linde:1984st}. Instead of restricting
ourselves to a massive scenario, a simple generalization is to
consider an arbitrary power index $p$, not necessarily fixed to
$p=2$~\cite{Silverstein:2008sg}. If the constant term is not zero,
then one obtains a small field
model~\cite{Linde:1981mu,Albrecht:1982wi} (with a negative second
term) or an effective hybrid model~\cite{Linde:1993cn,Copeland:1994vg}
(with a positive second term). Again, instead of considering only a
quadratic term, it is more generic to let the power index
unspecified. This leads to the three classes mentioned before.

\par

An important question is whether the other terms of the Taylor
expansion are under control. This has led to a debate on the question
of whether vacuum expectation values of $\phi$ larger than the Planck
mass are meaningful or
not~\cite{Lyth:1998xn,Linde:2005ht,Linde:2007fr}. In the simple
approach used here, we do not take part in this discussion and
consider sub- as well as super-Planckian vacuum expectation
values. Moreover, hybrid inflation is an intrinsic multiple field
scenario (with the above potential, inflation could not actually stop)
which cannot always be described by a single field
approach~\cite{Clesse:2009ur, Clesse:2010iz}. Indeed, in a multiple
field model, the presence of entropy perturbations can cause the
evolution of $\zeta_{\bmk}$ on large scales and this effect can modify
the power spectrum during the pre-heating stage. Since this type of
effect is model-dependent, it must be studied for each scenario and,
for this reason, it is wiser, in a first step, to focus on simpler
models. For this reason, we will consider in the following only the
large and small field scenarios having, respectively, the following
potentials:
\begin{equation} \label{eq:large_field}
  V(\phi)=M^4\left(\frac{\phi}{\Mp}\right)^p \text{ (large field),}
\end{equation}
and 
\begin{equation}\label{eq:small_field}
  V(\phi)=M^4\left[1-\left(\frac{\phi}{\mu}\right)^p\right] 
\text{ (small field)}.
\end{equation}
Of course, this has to be considered as a first step towards a more
complete scan of the inflationary landscape. The large field model is
characterized by two parameters, the energy scale $M$ and the power
index $p$. The small field potential is characterized by three
parameters, $M$, $\mu $ and $p$. We come back to the issue of the
prior distributions to assign to each parameter in
section~\ref{sec:priors}.

\subsection{Describing the Reheating}
\label{subsec:reheating}

In order to compare an inflationary model with observations, we also
need to take into account the reheating stage which takes place after
the end of inflation and before the onset of the radiation-dominated
era. This is compulsory since one needs to know the actual value of a
physical wavenumbers during inflation from its observed value
today. For instance, the amplitude of the power spectrum $P_*$ is
measured at a given wavenumber, typically $k_*/a_0=0.05 \Mpc^{-1}$,
where $a_0$ denotes the present-day scale factor. During inflation,
the corresponding physical wavenumber is stretched back to
\begin{equation}
\dfrac{k_*}{a} = \dfrac{k_*}{a_0} (1+z_\uend){\rm e}^{N_{\rm end}-N},
\end{equation}
where $z_\uend$ is the redshift at which inflation ended, $N_{\rm
  end}$ the total number of e-folds during inflation and $N\equiv \ln
a$ the number of e-folds at the time considered during inflation. The
quantity $k_*/a$ is uncertain precisely due to the existence of the
reheating. Assuming instantaneous transitions between inflation,
reheating, radiation and matter era, one can simplify
\begin{equation}
\label{eq:redend}
1+z_\uend = (1+z_\ueq) \left(\dfrac{\rhoreh}{\rhoeq} \right)^{1/4}
\dfrac{a_\ureh}{a_\uend}\,,
\end{equation}
where ``reh'' and ``eq'' respectively stands for the end of reheating
and the equality between the energy density of radiation and
matter. The so-called reheating parameter
$\Rrad$~\cite{Martin:2006rs,Martin:2010kz} describes the evolution of
the Universe during the reheating stage and is defined by
\begin{equation}
\label{eq:Rraddef}
\Rrad \equiv \dfrac{a_\uend}{a_\ureh} \left( \dfrac{\rhoend}{\rhoreh}
\right)^{1/4},
\end{equation}
such that Eq.~(\ref{eq:redend}) becomes
\begin{equation}
  1+z_\uend = \dfrac{1}{\Rrad} \left( \dfrac{\rhoend}{\rhoradnow} \right)^{1/4},
\end{equation}
where $\rhoradnow$ is the energy density of radiation today\footnote{The
  density parameter of radiation today is $\Omega_\radnow \simeq 2.471
  \times 10^{-5} h^2$.}. As a result, $\Rrad$ encodes all of our
ignorance on how the reheating influences the observable inflationary
power spectra. In fact, it is for inflation what the optical depth
$\tau$ is for CMB observations. The latter encodes how much
reionisation of the universe affects the measured CMB anisotropies
(independently of the details of the reionisation history, at least at
first order) while $\Rrad$ plays a similar role for the reheating. As
it should be clear from Eq.~(\ref{eq:Rraddef}), $\Rrad$ quantifies the
deviation from a reheating era which would be radiation-like.

\par

In fact, as discussed in Ref.~\cite{Martin:2010kz},
Eq.~(\ref{eq:Rraddef}) can be recast into various equivalent forms. In
terms of the number of e-folds during reheating $\Delta
N=N_\ureh - \Nend =\ln(a_\ureh/a_\uend)$, one has
\begin{equation}
\label{eq:Rrad}
\ln \Rrad = \frac{\Delta N}{4}\left(-1+3\overline{w}_{\ureh}\right),
\end{equation}
where $\overline{w}_{\ureh}$ stands for the mean equation of state
parameter
\begin{equation}
  \overline{w}_{\ureh}\equiv \frac{1}{\Delta N}\int _{\Nend}^{N_{\ureh}}
  \dfrac{P(n)}{\rho(n)}\, \ud n.
\end{equation}
Here $P(n)$ and $\rho(n)$ are the instantaneous total pressure and
energy density of the universe during reheating. This description is
completely general since no assumption about the physical properties
of the effective fluid dominating the matter content of the universe
during reheating has been made. One can also express $\Delta N$ in
terms of $\wreh$ such that
\begin{equation}
\label{eq:Rradw}
\ln \Rrad = \frac{1-3\overline{w}_{\ureh}}{12(1+\overline{w}_{\ureh})}\ln \left(
  \frac{\rhoreh}{\rhoend}\right).
\end{equation} 
As expected, one can verify explicitly that $\Rrad=1$ if
$\overline{w}_{\ureh}=1/3$.

\section{Bayesian Model Comparison}
\label{sec:bayesianevidence}

In this section, we briefly review Bayesian model comparison, which we
adopt to compare the performance of our inflationary models (for
further details, see \eg \cite{Trotta:2008qt}). As a preliminary
remark, we notice that if one seeks to determine the most economical
description of the inflationary potential in light of the available
data, Bayesian model comparison is well suited, in that
classical statistics only allows to reject hypotheses, not to confirm
them (see also Ref.~\cite{Liddle:2007fy} for alternative model selection
criteria). Therefore, while some simpler models might become ruled
out in a classical sense (\ie their parameter space can become
completely constrained by the data, until no viable region remains),
classical statistics does not allow one to rank the remaining models
in any way. Bayesian model comparison, with its natural inclusion of
the Occam's razor effect, is therefore the only available tool to
quantify in a self-consistent way our preference for a specific model.

\subsection{The Bayesian evidence}
\label{subsec:evid}

Bayesian model comparison aims at computing the posterior probability
of a model in view of the available data. The fundamental idea behind
the procedure is that ``economic'' models that fit well the data are
rewarded for their predictivity, while models with a large number of
free parameters that turn out not to be required by the data are
penalized for the wasted parameter space. Therefore, in a Bayesian
sense, the ``best'' model is the one that achieves the best compromise
between quality of fit and simplicity. One of the attractive features
of Bayesian model comparison is that it automatically embodies a
quantitative version of Occam's razor, i.e., the principle of
simplicity.

\par

Here and in the following, by ``model'' we denote a choice of
inflationary potential, together with a specification of its free
parameters, $\params_j$, {\em and} of their prior probability
distribution, $p(\params_j|\mdl_j)$. The specification of the prior is
fundamental for model comparison, as the prior shape and range
influence the Occam's razor effect.  From Bayes' theorem, the
posterior probability of model $\mdl_j$ given the data $d$,
$p(\mdl_j|d)$, is related to the Bayesian evidence (or model
likelihood) $p(d|\mdl_j)$ by
\begin{eqnarray} \label{eq:postM}
 p(\mdl_j|d)&=&\frac{p(d|\mdl_j)p(\mdl_j)}{p(d)}\, ,
\end{eqnarray}
where $p(\mdl_j)$ is the prior belief in model $\mdl_j$. In
Eq.~\eqref{eq:postM}, $p(d)=\sum_i p(d|\mdl_i)p(\mdl_i)$ is a
normalization constant (where the sum runs over all available known
models $\mdl_i$, $i=1,\dots, N$) and
\begin{equation} \label{eq:Bayesian_evidence}
  p(d|\mdl_j)=\int \dd
\params_j\, p(d|\params_j, \mdl_j) p(\params_j | \mdl_j)
\end{equation}
is the Bayesian evidence, where $p(d|\params_j, \mdl_j)$ is the
likelihood. The Bayesian evidence is thus the average likelihood under
the prior, and is the central quantity for Bayesian model comparison.

\par

Given two competing models, $\mdl_0$ and $\mdl_1$, the posterior odds
among them are given by \be \frac{p(\mdl_0 | d)}{p(\mdl_1 | d)} =
B_{01} \frac{p(\mdl_0)}{p(\mdl_1)} , \en where we have introduced the
factor $B_{01}$ as defined as the ratio of the models' evidences
\begin{eqnarray}
 B_{01}&\equiv&\frac{p(d|\mdl_0)}{p(d|\mdl_1)}\, .
\end{eqnarray}
The Bayes factor thus updates our relative state of belief in two
models from the prior odds to the posterior odds.  Large values of
$B_{01}$ denote a preference for $\mdl_0$, and small values of
$B_{01}$ denote a preference for $\mdl_1$. The ``Jeffreys' scale''
(Table~\ref{Tab:Jeff}) gives an empirical prescription for translating
the values of $B_{01}$ into strengths of belief.
\begin{table}[t]
 {\begin{tabular}{l l  l} \hline 
  $|\ln B_{01}|$ & Odds  & Strength of evidence \\\hline 
 $<1.0$ & $\lsim 3:1$ &  Inconclusive \\
 $1.0$ & $\sim 3:1$ &  Weak evidence \\
 $2.5$ & $\sim 12:1$ & Moderate evidence \\
 $5.0$ & $\sim 150:1$ &  Strong evidence \\
\hline
\end{tabular}}
\caption{Empirical scale for evaluating the strength of evidence when
  comparing two models, $\mdl_0$ versus $\mdl_1$ (so-called
  ``Jeffreys' scale'', here slightly modified following the
  prescriptions given in \cite{Gordon:2007xm,Trotta:2008qt}). The
  right-most column gives our convention for denoting the different
  levels of evidence above these thresholds.\label{Tab:Jeff} }
\end{table}

Given two or more models, specified in terms of their parametrization
{\em and} priors on the parameters, it is straightforward (although
sometimes computationally challenging) to compute the Bayes
factor. Depending on the problem at hand,
semi-analytical~\cite{Trotta:2005ar,Heavens:2007ka} and
numerical~\cite{Mukherjee:2005wg,Feroz:2007kg,Feroz:2008xx,
  Kilbinger:2009by,Serra:2007id} techniques are available. In the
usual case where the prior over models is taken to be non-committal
(\ie $p(\mdl_j) = 1/N$), the model with the largest Bayes factor
ought to be preferred. Thus the computation of $B_{01}$ allows to
select one (or a few) promising model(s) from a set of known
models. This framework has recently been extended to evaluate the
probability that the set of known models is incomplete, see
Ref.~\cite{March:2010ex}.

\par

Finally, we can also summarize our findings in terms of posterior
probability for the entire class of models being considered here,
large field or small field. From Bayes' theorem, the posterior
probability for \eg the small field class (SF) is given by 
\begin{equation} \label{eq:Post}
p(\text{SF} | \data) = \sum_{i=1}^{n_\text{SF}} \frac{p(\data |
  \text{SF}_i) p(\text{SF}_i)}{p(d)}, 
\end{equation} 
where 
\begin{equation} 
p(d) =
\sum_{i=1}^{n_\text{SF}} p(\data | \text{SF}_i) p(\text{SF}_i) +
\sum_{j=1}^{n_\text{LF}} p(\data | \text{LF}_j) p(\text{LF}_j) 
\end{equation} 
and $n_\text{SF} = 3$ is the number of small field models considered
in the class, while $n_\text{LF} = 6$ is the number of large field
models, as explained in the next section. Regarding the choice of
priors for the models, in view of comparing the viability of large
field and small field inflation, it is natural to divide equally the
prior probability between the two classes, and then further subdivide
it equally among the models in each class, so that $p(\text{SF}_j) =
1/(2 n_\text{SF})$ and $p(\text{LF}_j) = 1/(2 n_\text{LF})$. For
reasons that shall become clear below, it will be convenient to
consider the Bayes factor between the various models and the large
field model with $p=2$ (LF$_2$), and it is therefore useful to divide
both the numerator and the denominator of Eq.~\eqref{eq:Post} by the
evidence of LF$_2$, obtaining:
\begin{align} \label{eq:Post_2} p(\text{SF} | \data) & =
  \frac{\sum_i^{n_\text{SF}} B_{i*} p(\text{SF}_i)}{\sum_i^{n_\text{SF}}
    B_{i*} p(\text{SF}_i)
    + \sum_j^{n_\text{LF}} B_{j*} p(\text{LF}_j) } \\
  & =\frac{\langle B_{i*} \rangle_\text{SF}}{\langle B_{i*} \rangle_\text{SF} 
+ \langle B_{i*} \rangle_\text{LF}}  \\
  & =\left(1+ \frac{\langle B_{i*} \rangle_\text{LF}}{\langle B_{i*}
      \rangle_\text{SF}}\right)^{-1},
\end{align}
where we have defined 
\begin{align}
\langle B_{i*} \rangle_\text{SF} & 
\equiv \frac{1}{n_\text{SF}}\sum_{i=1}^{n_\text{SF}} B_{i*}, \label{eq:BavSF}\\
\langle B_{i*} \rangle_\text{LF} & 
\equiv \frac{1}{n_\text{LF}}\sum_{i=1}^{n_\text{LF}} B_{i*} \label{eq:BavLF}
\end{align}
and in the above $B_{i*}$ denotes the Bayes factor between model $i$
and the LF$_2$ model.
\par

\par

It is also instructive to consider the Bayesian complexity associated
with each model, defined as~\cite{Kullback:1951}
\begin{equation} \label{eq:Cb}
 \Cb = -2 \left[\DKL \left(P,\pi\right)- \widehat{\DKL} \right],
\end{equation}
where, here, $\pi$ denotes the prior distribution and
$\DKL\left(P,\pi\right)$ is the Kullback-Leiber divergence between the
posterior $P$ and the prior, $\pi$, namely
\begin{equation}
\DKL \left(P,\pi\right) \equiv \int p\left(\theta\vert d\right)
\log \frac{p\left(\theta\vert d\right)}{\pi(\theta)}{\rm d}\theta.
\end{equation}
In Eq.~\eqref{eq:Cb}, $\widehat{\DKL}$ denotes a point estimate for
the KL divergence. It has been shown
in~\cite{Kunz:2006mc,Trotta:2008qt} that the Bayesian complexity
measures the number of model parameters that the data can
constrain. Evaluated together with the evidence, the complexity helps
to assess whether the parametrization of a model is excessive for the
constraining power of the available data (for details,
see~\cite{Kunz:2006mc}). The complexity can be expressed
as
\begin{equation}
\label{eq:complexity_chisq}
\DKL = \langle \chi^2 \rangle -
\widehat{\chi}^2,
\end{equation}
where $\chi^2 \equiv -2 \ln \like$ and the
expectation value is taken with respect to the posterior. The second
term, $\widehat{\chi}^2$ is a plug-in estimate that can be taken to be
for example the best-fit $\chi^2$ value or the value of the $\chi^2$
at the posterior mean. Here we adopt the best-fit value,
following~\cite{Kunz:2006mc}.

\par

As mentioned above, the evidence is computed using the publicly
available {\tt MultiNest}
code~\citep{Feroz:2007kg,Feroz:2008xx,Trotta:2008bp}, which implements
the nested sampling algorithm.  The gist of nested sampling is that
the multi-dimensional evidence integral of
Eq.~\eqref{eq:Bayesian_evidence} is recast into a one-dimensional
integral. This is accomplished by defining the prior volume $x$ as
${\rm d} x \equiv p(\params_j | \mdl_j){\rm d}
\params_j $ so that
 \begin{equation} \label{eq:def_prior_volume}
  x(\lambda) = \int_{\like(\params_j)>\lambda} p(\params_j | \mdl_j) {\rm d}
  \params_j,
 \end{equation}
 where the integral is over the parameter space enclosed by the
 iso-likelihood contour $\like(\params_j) = \lambda$. So $x(\lambda)$
 gives the volume of parameter space above a certain level $\lambda$
 of the likelihood (for a specific model $\mdl_j$). Then the Bayesian evidence,
 Eq.~\eqref{eq:Bayesian_evidence}, can be written as
 \begin{equation} \label{eq:nested_integral}
 p(\data | \mdl_j) = \int_0^1 \like(x) {\rm d} x,
 \end{equation}
 where $\like(x)$ is the inverse of
 Eq.~\eqref{eq:def_prior_volume}. Samples from $\like(x)$ can be
 obtained by drawing uniformly samples from the likelihood volume
 within the iso-contour surface defined by $\lambda$. The standard
 deviation on the value of the log evidence can be estimated as
 $(H/n_\text{live})^{1/2}$, where $H$ is the negative relative entropy
 and $n_\text{live}$ is the number of live points adopted, which in
 our case is $n_\text{live} = 1000$ (see~Ref.~\cite{Feroz:2007kg} for
 details). We have checked that our evidence values are robust (within
 error bars) if one increases $n_\text{live}$ to $5000$. The posterior
 distributions have also been cross-checked with standard
 Metropolis--Hastings Markov--Chain--Monte--Carlo (MCMC).

\subsection{Choice of Priors}
\label{sec:priors}

\begin{table*}
  \begin{tabular*}{0.85\textwidth}{@{\extracolsep{\fill}}| l | c c c | c c c c c c | } \hline
    Parameter & \multicolumn{3}{c|}{Small field models, Eq.~\eqref{eq:small_field}}  & \multicolumn{6}{c|}{Large field models,  Eq.~\eqref{eq:large_field} } \\
    & SFI$_s$ & SFI$_l$  & SFI$_f$ & LFI$_p$ &   LFI$_{2/3}$ & LFI$_1$ &  LFI$_2$ &  LFI$_3$ &  LFI$_4$ \\ \hline
    Normalization, $\ln P_*$ & \multicolumn{3}{c|}{$[2.7\times 10^{-10}, 4.0\times 10^{-10}]$} & \multicolumn{6}{c|}{$[2.7\times 10^{-10}, 4.0\times 10^{-10}]$}   \\ 
    Exponent, $p$ & \multicolumn{3}{c|}{$[2.4, 10]$} & $[0.2, 5]$ & $2/3$ & $1$& $2$ & $3$ & $4$\\
    Vacuum expectation, $\log(\mu/\Mp)$ & $[-1,0]$ & $[0,2]$ & $[-1,2]$ & \multicolumn{6}{c|}{Not applicable}\\ 
    Reheating,  $\ln R$ & \multicolumn{3}{c|}{$[-46, 15]$}  & \multicolumn{6}{c|}{$[-46, 15]$}  \\ \hline
    $n$ &  4 & 4 & 4 & 3 & 2 & 2 & 2 & 2 & 2 \\

        \hline
\end{tabular*}
\caption{Inflationary models considered in this analysis and priors on their parameters. All priors are taken to 
  be uniform (\ie flat) in the variable and range specified, see the text 
  for a detailed justification. In the last row, $n$ is the number of free 
  parameters related to the inflationary sector. }
  \label{tab:models}
\end{table*}

Since our aim is to evaluate the evidence of large and small field
models, it is absolutely crucial to choose well-motivated priors for
the parameters describing the potential. In order to see why it is so,
it is instructive to consider the evidence of a simple, one-parameter
toy case, where there is only one single parameter $\theta$, whose
prior density under model $\mdl$ is given by $p(\theta | \mdl)$. We
shall further assume that the likelihood is much more sharply peaked
than the prior (\ie the quantity $\theta$ has been well measured), so
that $p(\theta) \approx \text{const.}$ in the range $\delta\theta$
where the likelihood $\like(\theta)$ is appreciably different from
zero. Then the evidence of model $\mdl$,
Eq.~\eqref{eq:Bayesian_evidence}, is approximately equal to
\begin{equation}
\label{eq:penalty}
p(\data | \mdl ) \approx
\like(\theta_\text{ML}) \, \delta\theta \, p(\theta_\text{ML} | \mdl),
\end{equation}
where $\theta_\text{ML}$ is the value that maximizes the likelihood
function. Since the prior must be normalized, $p(\theta_\text{ML} |
\mdl) \approx 1/\Sigma$, where $\Sigma$ is the characteristic width of
the prior. Therefore one finds that $p(\data | \mdl ) \propto
\Sigma^{-1}$, \ie the evidence scales inversely proportionally to the
width of the prior. The term $\delta\theta/\Sigma$ is the so-called
``Occam's factor'', which penalizes models with a large ``wasted''
parameter space under the prior, \ie models for which the
characteristic width of the likelihood is much smaller than that of
the prior, $\delta\theta/\Sigma \ll 1$. Hence the \emph{a priori}
plausible range of parameter values determines the strength of the
Occam's penalty term, and for this reason it has to be carefully
chosen on the basis of physical considerations\footnote{Notice that
  parameters which are unconstrained by the data are not penalized by
  the Occam's factor, \ie if the likelihood's width is similar to the
  prior range, then $\delta\theta/\Sigma \sim 1$ and the Occam's
  factor effect vanishes.}.

\par
 
Going back to the potentials \eqref{eq:large_field} and
\eqref{eq:small_field}, we notice that the parameter $M$, common to
both classes of models, is {\em a priori} unknown, and is
observationally determined by the overall normalization of the power
spectrum, $P_*$. Since the \emph{scale} of $M$ is unknown, it is
appropriate to adopt a prior flat on $\ln M$, to reflect the fact that
we are giving equal a priori probability to all orders of magnitude
within some suitably chosen lower and upper limits. A flat prior on
$\ln M$ is equivalent to a flat prior on $\ln P_*$, and therefore in
our numerical sampling we swap $\ln M$ for $\ln P_*$ as a fundamental
parameter. Since the overall power spectrum normalization is common to
all models, the precise range of values under the prior for $\ln P_*$
becomes irrelevant (as long as the range is sufficiently wide to
encompass the support of the likelihood), as all models share the same
Occam's razor penalty from this common parameter. In practice, we
chose $\ln P_* \in [2.7 \times 10^{-10}, 4.0 \times 10^{-10}]$, but
because of the above argument the Bayes factor between our models
would remain unchanged even if this range was arbitrarily enlarged.

\par

For large field models, we chose to adopt a flat prior in the range
$0.2<p<5$. The lower limit is arbitrarily chosen to encompass all
proposed large field potentials having a fractional
power~\cite{McAllister:2008hb,Kallosh:2010ug}. In principle, one could
imagine an arbitrarily small $p$ (which would suggest the use of a
Jeffreys' prior, instead) but, up to now, there is no theoretical
motivation to do so. On the other hand, there is no strong theoretical
reason not to consider a model with, say, $p=7$. However, we know that
the data already strongly disfavour models with $p>5$ (as a matter of
fact, even models with $p>3$ are disfavoured~\cite{Martin:2010kz}) and
therefore one expects that the evidence of models with $p>5$ (fixed)
would be strongly disfavoured. Furthermore, if one wanted to enlarge
the prior range to $p>5$ it would be easy to rescale the evidence to
account for the enlarged parameter space, since the likelihood is
close to 0 for $p>5$. This would lead to a larger Occam's penalty and
thus to a lower evidence, see Eq.~\eqref{eq:penalty}.

\par

For small field models, we have chosen a flat prior $p\in [2.4,10]$
as our representative class since $p=2$ is a very special
case. As discussed in Ref.~\cite{Martin:2006rs}, approaching the value
$p=2$ is numerically tricky and we have chosen the lower bound as the
closest, but different, possible value of $p>2$. Models with
$p<2$~\cite{Alabidi:2005qi, Alabidi:2008ej} might in principle be
included but would constitute another class of models since this would
require to cross the $p=2$ barrier. Moreover, models with negative
$p$ correspond to very different physical regimes. For instance, the
model with $p=-4$ is nothing but the Coulomb potential of brane
inflation and was analyzed in detail in~\cite{Lorenz:2007ze}. For the
reasons detailled in Sec.~\ref{sec:introduction}, and at this stage of
the analysis, we do not include those cases. The upper bound
for $p$ has been chosen typically an order of magnitude
higher. Theoretically, as already mentioned above, small field models
are archetypal of inflationary potentials which can be Taylor expanded
in the (small) field values, in units of a given vacuum expectation
value $\mu$. As a result, too large values of $p$ would appear quite
unnatural. Concerning $\mu$, its scale is a priori unknown and,
therefore, we have chosen a flat prior on $\log\left(\mu/\Mp\right)$
in the range $[-1,2]$. On one hand, if one has a theoretical prejudice
of viewing the small field models as representative of Taylor expanded
potential (as was done above), and in particular in the supersymmetric
framework, one would expect $\mu < \Mp$ to keep the supergravity
corrections under control. On the other hand, other theoretical
approaches do not forbid super-Planckian vacuum expectation
values~\cite{Brax:2005jv} since one can always consider that this
potential is obtained, not from a Taylor expansion but exactly from a
more fundamental theory. The corrections would therefore not be
controlled by the ratio of the vacuum expectation value to the Planck
mass but by the ratio of the energy density to the Planck
density. Hence our prior range is chosen in such a way as to extend
above the Planck mass. Concerning the boundary values, in the limit
$\mu/\Mp\gg 1$, one can show that the two first slow-roll parameters,
and hence all observable predictions, do not longer depend on both
$\mu$ and $p$. As a result, it is straightforward to show from
Eq.~(\ref{eq:penalty}) that the corresponding Bayes factors would be
unchanged for larger values of $\mu$. In the limit $\mu/\Mp\ll 1$,
the first slow-roll parameter becomes tiny and the second one becomes
$\mu$-independent such that, again, the observable predictions, and
thus the likelihood and the evidence, are no longer sensitive to
$\mu$.

\par

Finally, in addition to the two broad classes of large field and small
field models, we have introduced in our model space finer subdivisions
leading to more specific model classes. Motivated by the above prior
discussion, it is natural to further distinguish between small field
models allowing super-Planckian expectation values [\ie with $\log
(\mu/\Mp) > 0$] from the ones that do not [$\log(\mu/\Mp) < 0$]. In the
large field class, we have also singled out some models having a
peculiar interest such as the genuine chaotic massive inflation model
($p=2$), monodromy inflation ($p=2/3$), linear inflation ($p=1$), and
the self-interacting potential ($p=3$ or $p=4$).  Of course, one must
restrict oneself to the positive part of the potential when
necessary. Therefore we consider a total of 9 classes of models.

\par

Having parametrized the evolution of the universe during the
reheating in the previous section, one must now discuss the choice of
the prior on the reheating parameter. As shown in
Ref.~\cite{Martin:2006rs}, instead of working with $\Rrad$ introduced
in Eq.~\eqref{eq:Rraddef}, it is more convenient to work with the
rescaled reheating parameter $R$ defined by
\begin{equation}
\label{eq:Rdef}
R\equiv \Rrad\frac{\rhoend^{1/4}}{\Mp}\,.
\end{equation}
As we recap in the appendix~\ref{sec:R}, $\Rrad$ exhibits trivial
correlations with the normalisation of the power spectrum $P_*$ which
can be easily removed by considering $R$ instead. Notice that once the
inflationary model is specified, $\rhoend$ is known and
Eq.~(\ref{eq:Rdef}) is nothing but a rescaling. Clearly, the order of
magnitude of the different physical quantities appearing in
Eqs.~(\ref{eq:Rraddef}) and (\ref{eq:Rdef}) is unknown and this
suggests that we choose a flat prior on $\ln R$. The next step is to
determine the prior boundaries. In fact, using the expression of
$\Rrad$ given before, one also has
\begin{equation}
\ln R=\frac{1-3 \wreh}{12(1+\wreh)}\ln \left(
\frac{\rhoreh}{\Mp^4}\right)+\frac{1+3 \wreh}
{6(1+ \wreh)}\ln 
\left(\frac{\rhoend}{\Mp^4}\right).
\end{equation} 
Positivity energy conditions in General Relativity imposes that
$\wreh$ cannot exceed unity and we want to separate inflation from
reheating such that $\wreh$ cannot be less than $-1/3$. Moreover,
$\rhonuc<\rhoreh<\rhoend$, where $\rhonuc$ is the energy density at
Big-Bang Nucleosynthesis (BBN), which we take to be $\rhonuc^{1/4} =
10\, \MeV$, this implies that $-46<\ln R<15 + (1/3)
\ln\left(\rhoend/\Mp^4\right)$. Since there is no preferred value for
$\ln R$, we initially take the maximal possible theoretically
allowed range $[-46,15]$. However, for each given model parameter
values, we then reject all $\ln R$ values not satisfying the
consistency bound $\ln R<15 + (1/3)
\ln\left(\rhoend/\Mp^4\right)$. Finally, notice that this description
of reheating via the $\ln R$ parameter and its prior range is common
to all models.

\par

To conclude the discussion on priors, we have chosen flat priors on
the standard cosmological parameters centered around their currently
measured values, \ie for the density parameter of baryons $\OmegaB
h^2$, of dark matter $\OmegaDM h^2$, the angular size of the sound
horizon at last scattering $\theta$ and the optical depth $\tau$. We
also marginalize over the amplitude of the unresolved SZ signal with a
flat prior in the range $A_\text{SZ} \in [0, 2]$. These prior choices
do not impact on our evidence result for the inflationary models as
all models share the same standard cosmological parameters and their
respective priors. We moreover assume throughout a flat universe as
predicted by cosmic inflation.

\par

The models we consider and the priors on the relevant inflationary
parameters are summarized in Table~\ref{tab:models}.

\section{Results and Discussion}
\label{sec:discussion}

\begin{figure*}
\begin{center}
\includegraphics[width=0.95\textwidth,clip=true]{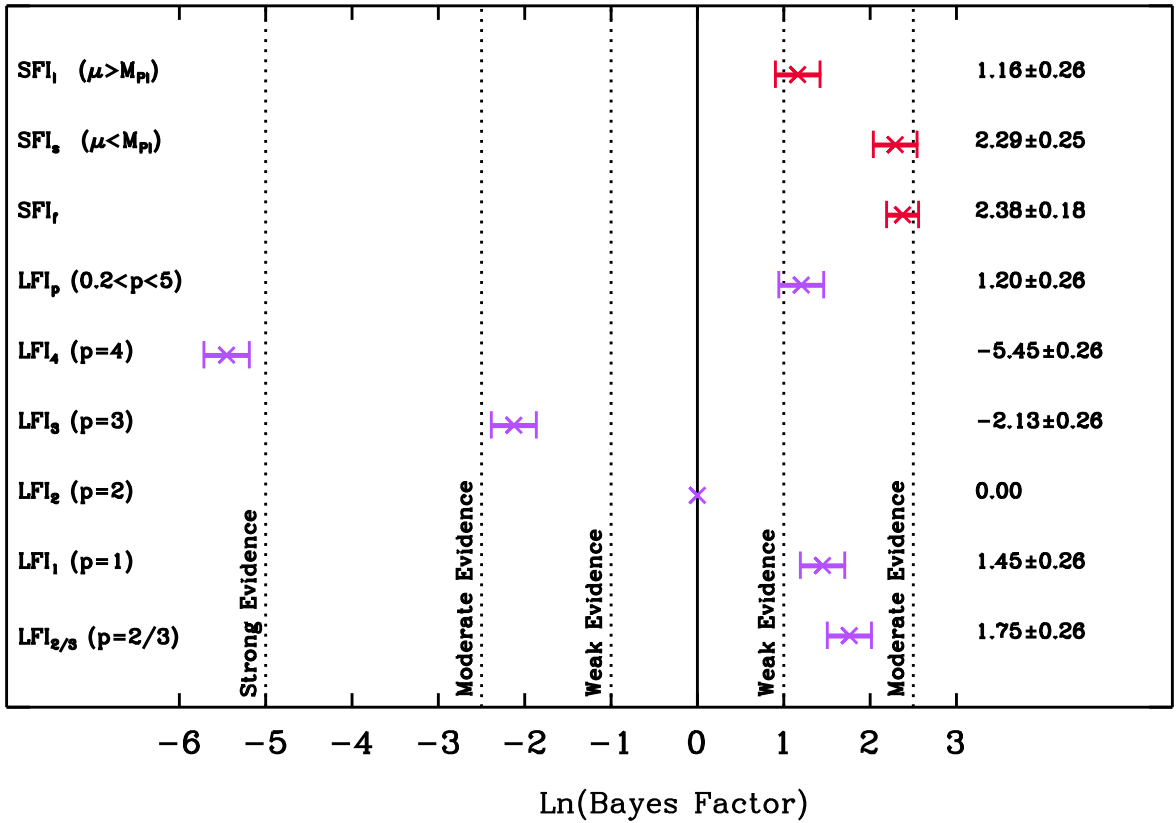}
\caption{Results for the Bayes factor between different inflationary
  models considered in the analysis. The names of the models are
  specified on the left of the figure. The Bayes factor are computed
  taking massive large field model as the reference model, and the
  results are given in the column on the right of the plot. The dotted
  vertical lines indicate the thresholds of weak, moderate and strong
  evidence, as per Table~\ref{Tab:Jeff}.}
\label{fig:evidence}
\end{center}
\end{figure*}

In this section, we present our model comparison results for the
classes of models described above. Concerning the data, we have used
the seven years Wilkinson Microwave Anisotropies Probe (WMAP7)
data~\cite{Komatsu:2010fb, Larson:2010gs, Jarosik:2010iu} complemented
with the Hubble Space Telescope (HST) constraints on the Hubble
constant today, $H_0 = 74.2 \pm 3.6$ km/s/Mpc~\cite{Riess:2009pu}. Our
findings are summarized in Fig.~\ref{fig:evidence}, where we show the
Bayes factors for each model, computed with respect to the large field
model with $p=2$.

\par

Within the class of large field models, we can see that models with $p
\geq 3$ are disfavoured, at the ``weak evidence level'' for $p=3$ and
at the ``strong evidence'' level for $p=4$. Clearly, one can conclude
that models with even larger (and fixed) values of $p$ would be even
more strongly disfavoured, so that they can be effectively ruled
out. We have chosen the large field $p=2$ model as our ``reference
model'' (the one with respect to which the Bayes factors are computed)
because it plays the role of a watershed point: large field models
with shallower potentials are preferred by the Bayesian evidence, with
$p=1$ and $p=2/3$ gathering slightly more than ``weak evidence'' in
their favour. However, the evidence is not strong enough to allow one
to conclude a definite preference for these models. The more generic
large field model with $p\in [0.2,5]$ is also weakly preferred over
LF$_2$, and this despite the extra parameter of the former, which
incurs an Occam's razor penalty. As expected, the performance of this
model, as measured by the evidence, falls in between the steep
potentials ($p>2$) and the shallower ones ($p<2$).

\par

Moving on to small field models, we remark that their overall
performance is superior to our reference large field model LF$_2$, but
quite comparable to the shallower large field models, despite the fact
that small field models have one or even two parameters more than
large field models. The very best models of inflation are small
field. Within the error bars, the evidence cannot distinguish between
a model with an upper cutoff at $\Mp$ and one that allows $\mu$ to go
above the Planck mass. Models with purely super-Planckian expectation
values are only very slightly disfavoured, by about 1 unit in the
log evidence. Therefore we can conclude that the data are presently
not sufficient to distinguish between the two scenarios.

\par
\begin{figure}
\begin{center}
\includegraphics[width=0.5\textwidth,clip=true]{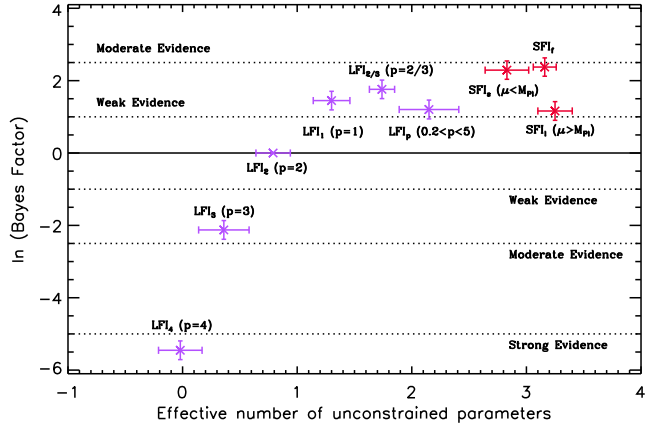}
\caption{Bayes factor versus the effective number of unconstrained
  parameters ($\Delta \Cb$) for all large and small field models. The
  steeper LFI models are genuinely disfavoured by the data, as all of
  their free parameters are well constrained. Small field models being
  favoured by the evidence still have unconstrained parameters, and
  therefore it is too early to conclusively rule out shallower ($p<2$)
  large field models, despite the fact that they exhibit a slightly
  smaller Bayes factor. }
\label{fig:complexity}
\end{center}
\end{figure}

 
Further insight in the model comparison outcome can be garnered by
investigating simultaneously the Bayesian complexity and the evidence
(or the Bayes factor) of the models considered here (see
Ref.~\cite{Kunz:2006mc} for further details about the interpretation
of the complexity). The Bayesian complexity,
Eq.~\eqref{eq:complexity_chisq}, has been computed for each model from
a pure MCMC run whose convergence has been monitored by using the
R statistics implemented in \texttt{CosmoMC}~\citep{Lewis:2002ah}. The
chains have been stopped as soon as the estimated errors were below
$3\%$, which corresponds to a total number of samples ranging from
$5\times 10^4$ to $4\times 10^5$ depending on the underlying
inflationary model. The variance of our complexity estimate is
obtained from the variance of four sub-chains of equal length randomly
selected from the post burn-in samples. Both quantities are displayed
in Fig.~\ref{fig:complexity}, where the horizontal axis gives the
value of the number of input parameters for each model (both
inflationary and cosmological) minus the Bayesian complexity, which we
denote by the symbol $\Delta\Cb$. A value of $\Delta\Cb$ close to zero
means that the model parameters are well constrained by the data,
while $\Delta\Cb > 0$ gives an estimate of the effective number of
parameters remaining unconstrained by the data.

The value of $\Delta \Cb$ for the large field models with $p>2$ is
generally smaller, and reaches $\Delta\Cb \approx 0$ for $p=4$, the
model with the lowest evidence. This is a consequence of the tension
between these models and the data, which leads to the reheating
parameter becoming more and more constrained as $p$ increases: for
$p=4$, we find a 2$\sigma$ lower limit $\ln \Rreh > -2.1$, thus
leading to an increase in the value of the complexity by about 1
unit. Since the models with $p=3$ and $p=4$ have the smallest Bayes
factor while exhibiting values of $\Delta\Cb$ close to 0 (meaning that
all of their free parameters are well constrained), we can conclude
that those models are genuinely disfavoured by the data. On the other
hand, for the models having a similar Bayes factor,
Fig.~\ref{fig:complexity} shows that the larger number of free
parameters in the small field models corresponds to an increase in the
number of unconstrained parameters $\Delta\Cb$ with respect to its
value for the large field models with $p\leq 1$. This indicates that
the extra inflationary parameters in the small field class are not
being constrained by the data. Therefore we are led to conclude that
while a slight preference for small field models is beginning to
accumulate, it is too early to be able to conclusively favour small
field models over large field ones. It is expected that Planck data
will be able to conclusively pass judgement on this issue.

A consistent picture emerges when one considers the Bayesian
complexity of the two models with the largest number of parameters in
each class, namely SFI$_f$ and LFI$_p$, with 4 and 3 inflationary
parameters, respectively. For both cases, we find a similar
complexity, $\Cb \simeq 5.9$, which suggests that current data can
constrain up to approximately $2$ inflationary parameters. This is
because our models have all $N=5$ non-inflationary parameters in
common, including the SZ amplitude, and 4 of them are well constrained
and contribute approximately 4 units to the Bayesian complexity. This
leads to the conclusion that WMAP7 data are still insufficiently
powerful to fully constrain the whole inflationary sector as
parametrized in this work (see also Refs.~\cite{Kawasaki:2009yn,
  Parkinson:2010zr}).

\par 

We can also evaluate the posterior probability for the entire class of
small field scenarios. From Eq.~\eqref{eq:Post_2} we find
\begin{equation}
p\left({\rm SF}\vert d\right)\simeq  0.77\pm 0.03,
\end{equation}
and, therefore, $p\left({\rm LF}\vert d\right)\simeq 0.23 \pm
0.03$. Therefore, the probability of the small field scenario has
risen from 50\% in the prior to 77\% in the posterior. This represents
posterior odds of $\sim 3:1$ in favour of small field inflation, as
compared with large field inflation. Although, as explained above,
this shift in the odds is by no means conclusive, it does represent an
indication that large field inflation is getting increasingly under
pressure from the data~\cite{Boyanovsky:2009xh}.

Finally, it is important to assess the robustness of our results with
respect to reasonable changes in our choice of models' priors. Our
choice to divide the prior probability equally between the LF class
and the SF class reflects the desire to compare both classes of models
on an equal footing \emph{a priori}. Another natural choice for the
models' prior would be to split the prior mass equally among models,
i.e. to assign $p({\rm SF}_i) = p({\rm LF}_i) = 1/(n_{\rm SF} + n_{\rm
  LF})$.  This choice would however result in prior odds of 2:1 in
favour of the LF class, which seems contrived, given that it arises
solely from the fact that we have double as many LF models as SF
models. Even with this (unfair to the SF class) prior choice, the
posterior probability for SF would be $p({\rm SF} | d) \simeq 0.6$ [up
from an initial prior probability $p({\rm SF}) = 1/3$], so our result
of a (slight) preference for SF models stands.

Finally, we notice that our result is robust with respect to the
inclusion of further models under either the SF or LF class, provided
such models are disfavoured by the data (as they would be e.g.~for
$p>5$ in the LF class). Inclusion of such highly disfavoured models
would result in their Bayes factors with respect to $\rm LF_2$ being
close to $0$, hence the average values defined in
Eqs.~\eqref{eq:BavSF} and ~\eqref{eq:BavLF} would simply be rescaled
by the new (larger) number of models in each class. However, the
posterior probability of SF models only depends on the ratio of the
average Bayes factors [see Eq.~\eqref{eq:Post_2}], hence such
rescaling factors would largely cancel out (for a detailed discussion
of this rearrangement of prior probability in a similar context, see
Ref.~\cite{March:2010ex}). This holds true provided the overall number
of models in each class is not widely different. We do not have any
reason to believe that this should be the case. However, if one of
the model classes truly had a much larger number of potential models
in it, one would have to carefully reconsider the choice of giving
both classes equal \emph{a priori} mass: after all, a class of models
with a smaller number of physically distinct possibilities in it is
\emph{a priori} more predictive than a class with a large number of
possible distinct models.

\section{Conclusion}

To summarize, this article presented the first calculation of the
Bayesian evidence for different classes of inflationary scenarios,
explaining from first principles how physically meaningful priors
could be derived for the fundamental parameters of the models. Among
the models studied here, small field models appear to be favoured,
albeit still in a fairly mild way. This result must be viewed as a
first step towards a more exhaustive exploration of the inflationary
landscape. With the techniques introduced here and the high accuracy
CMB data soon available, we have paved the way to the identification
of the best inflationary scenario.

\begin{acknowledgments}
  We would like to thank Patrick Peter and Jean-Philippe Uzan for
  useful discussions. RT would like to thank the Office of the Mayor
  of the City of Paris for partial support and the Institut
  d'Astrophysique de Paris (IAP) for hospitality. This work is
  partially supported by the Belgian Federal Office for Science,
  Technical and Cultural Affairs, under the Inter-university
  Attraction Pole Grant No. P6/11
\end{acknowledgments}

\appendix*

\section{Optimal reheating parameter}
\label{sec:R}

As discussed in Ref.~\cite{Martin:2010kz}, the reheating parameter
$\Rrad$ can also be explicitly related to observable quantities and to
the number of e-folds $N_*$ at which a given scale $k_*$ leaves the
Hubble radius, i.e. when $k_* = a(N_*) H(N_*)$ during inflation
\begin{equation}
\begin{aligned}
\ln \Rrad \simeq (\Nend - \Nstar) +N_0
& -\frac14\ln\left(8 \pi^2 P_*\right)
 \\ & +
\frac{1}{4}\ln\left( \dfrac{72}{r} \dfrac{V_\uend}{V_*} \right),
\end{aligned}
\label{eq:Rradobs}
\end{equation}
where $V_*$ stands for the potential evaluated at the e-fold $N_*$,
\ie when $\phi=\phi(N_*)$. The quantity $r$ is the primordial
tensor-to-scalar ratio and
\begin{equation}
N_0\equiv \ln\left(\dfrac{k_*/a_0}{\rhoradnow^{1/4}} \right).
\end{equation}
Using the Friedmann--Lema\^{\i}tre equations together with
Eq.~(\ref{eq:Rdef}), the rescaled reheating parameter now reads
\begin{equation}
\label{eq:Rrehsr}
\begin{aligned}
  \ln R & \simeq (\Nend - \Nstar) + \Nzero + 
\dfrac{1}{2} \ln\left(\dfrac{9}{2}
    \dfrac{V_\uend}{V_*} \right), 
\end{aligned}
\end{equation}
which clearly no longer depends explicitly on $P_*$. There is thus a
great advantage of sampling the model parameters on $\ln R$ rather
than $\ln \Rrad$ to prevent the unwanted degeneracies appearing in
Eq.~(\ref{eq:Rradobs}). This was the approach adopted recently in
Ref.~\cite{Martin:2010kz} where, in most of this work, the parameter
$\ln R$ was used. Additional constraints can be further obtained if
one introduces extra assumptions on the equation of state
parameter. This question was also studied in
Ref.~\cite{Martin:2010kz}, but only for specific cases (unlike the
description of that work given in Ref.~\cite{Mortonson:2010er}). Let
us notice that in Ref.~\cite{Mortonson:2010er}, the reheating phase
was marginalised over by using a prior on $\Delta N_* =\Nend -N_*$ for
a particular value of $k_*=0.05\,\Mpc^{-1}$. From the above formula,
it is clear that one can always trade the parameter $\ln R$ with
$\Delta N_*$ provided some values for $r$ and $P_*$ are assumed,
although it might seem awkward to introduce a scale dependent prior
for a background quantity (the prior changes if one chooses another
value for $k_*$). A drawback of this approach is that this introduces
correlations between the parametrization of reheating and the
normalization of the power spectrum when one has to determine the
prior range. Reference~\cite{Mortonson:2010er} fixes $20<\Delta
N_*<\Delta N_*^{\uend}$, where $\Delta N_*^{\uend}$ corresponds to the
value of $\Delta N_*$ when $\rhoend=\rhoreh$. It is easy to see that
this choice excludes from the prior models that are physically
legitimate. For instance, a small field model with $p=3$, $\wreh=-0.3$
and $\mu=0.01\Mp$ is such that $17.2<\Delta N_*<46.0$ (for this model,
the lower bound is always smaller than $20$ for $\mu
\in[0.01,10]$). Of course, this also depends on the choice of
$\rhonuc$ which is not given very precisely. If one takes
$\rhonuc\simeq \left(100\, \MeV \right)^4$, the lower bounds become
$\simeq 19.3$ but if one chooses the extreme value $\rhonuc\simeq
\left(1\, \MeV \right)^4$, it is $\simeq 15.1$. When constraining
model parameters, these considerations does not really matter since,
as shown in Ref.~\cite{Martin:2010kz}, these model parameters are in
fact disfavoured by the data. But, clearly, this will affect the
calculation of the evidence. The above considerations show that priors
should always be chosen and justified from physical considerations.

\par

Finally, in this article we have restricted ourselves to a standard
post-reheating thermal history. But the approach used here can in fact
be straightforwardly generalised to a non-standard thermal history
before Big-Bang Nucleosynthesis (BBN). For instance, if one assumes
that, inserted into the radiation-dominated era, there is actually a
phase of evolution dominated by a fluid $\mathrm{X}$ the equation of
state parameter of which is given by $w_{_\mathrm{X}}$, one could
define a new parameter $R_{_\mathrm{X}}$ by
\begin{equation}
\ln R_{_\mathrm{X}}\equiv \frac{1-3\overline{w}_{_\mathrm{X}}}
{12\left(1+\overline{w}_{_\mathrm{X}}\right)}
\ln \left(\frac{\rho_{_\mathrm{X}}^{\rm end}}{\rho_{_\mathrm{X}}^{\rm start}}
\right),
\end{equation}
where $\rho_{_\mathrm{X}}^{\rm start}$ is the energy density at the
beginning of the epoch dominated by the fluid $\mathrm{X}$ and
$\rho_{_\mathrm{X}}^{\rm end}$ the energy density at the end. Then,
nothing is changed in the above description except that the parameter
$\Rrad$ should now be replaced with $\Rrad R_{_\mathrm{X}}$. Moreover,
if after the $\mathrm{X}$-dominated period, there is another, say,
$\mathrm{Y}$-dominated period, then one can define the parameter
$R_{_\mathrm{Y}}$ and $R_{\rm rad}$ should now be replaced with $\Rrad
R_{_\mathrm{X}} R_{_\mathrm{Y}}$. Obviously, this works for an
arbitrary number of new epochs. This also means that these
non-standard thermal histories are not really observable (unless one
has a definite model for the reheating) since the new parameters
$R_{_\mathrm{X}}$ and $R_{_\mathrm{Y}}$ are in fact completely
degenerated with $\Rrad$.

\bibliography{biblio}

\end{document}